\documentclass{article}

\usepackage{PRIMEarxiv}

\usepackage[utf8]{inputenc} 
\usepackage[T1]{fontenc}    
\usepackage{hyperref}       
\usepackage{url}            
\usepackage{booktabs}       
\usepackage{amsfonts}       
\usepackage{nicefrac}       
\usepackage{microtype}      
\usepackage{lipsum}
\usepackage{fancyhdr}       
\usepackage{graphicx}       
\usepackage{amsmath,amssymb}
\graphicspath{{media/}}     

\newcommand{\bbR}{\mathbb{R}}

\title{Physics-Inspired Unsupervised Classification for Region of Interest in X-Ray Ptychography
}

\author{
  Dergan Lin \\
  Mathematics and Computer Science Division \\
  Argonne National Laboratory \\
  Lemont, IL 60439, USA\\
  \texttt{dlin@anl.gov} \\
   \And
  Yi Jiang \\
  Advanced Photon Source \\
  Argonne National Laboratory \\
  Lemont, IL 60439, USA\\
  \texttt{yjiang@anl.gov} \\
     \And
  Junjing Deng \\
  Advanced Photon Source \\
  Argonne National Laboratory \\
  Lemont, IL 60439, USA\\
  \texttt{junjingdeng@anl.gov} \\
     \And
  Zichao Wendy Di \\
  Mathematics and Computer Science Division \\
  Advanced Photon Source \\
  Argonne National Laboratory \\
  Lemont, IL 60439, USA\\
  \texttt{wendydi@mcs.anl.gov} \\  
}

\begin{document}
\maketitle

\begin{abstract}
X-ray ptychography allows for large fields to be imaged at high resolution at the cost of additional computational expense due to the large volume of data. 
Given limited information regarding the object, the acquired data often has an excessive amount of information that is outside the region of interest (RoI).
In this work we propose a physics-inspired unsupervised learning algorithm to identify the RoI of an object using only diffraction patterns from a ptychography dataset before committing computational resources to reconstruction. 
Obtained diffraction patterns that are automatically identified as not within the RoI are filtered out, allowing efficient reconstruction by focusing only on important data within the RoI while preserving image quality.
\end{abstract}


\section{Introduction}
X-ray transmission ptychography is an imaging technique that can achieve nanoscale resolution through measuring multiple diffraction patterns by scanning an X-ray probe across an object of interest~\cite{pfeiffer2018x, suzuki2014high, thibault2008high}.
Typical ptychography experiments collect a large number of diffraction patterns to cover an extended region with sufficient overlapping to achieve high-resolution reconstruction images.
However, traditional 2D scans used to capture the object usually include a large portion of unwanted or unnecessary regions, for example, empty background, thus imposing a high expenditure of computational resources on the reconstruction of regions that may not be of interest.
Currently in X-ray ptychography, in order to reduce the data acquired from an area that is not a region of interest (RoI), an initial coarse scan on a large field of view (FoV) is performed to gain a rough idea of where the object's RoI could be, this is then followed by a finer scan on a smaller FoV. While this approach reduces the scanning region, a rectangular or square scan FoV can still cover a significant portion of unwanted data points, especially when the scanned object is particle or cell type. In some cases, such as ptychographic tomography, additional empty regions are even added on purpose to increase the scan FoV to accommodate sample drifts or rotation errors, which are expected to be uninformative for the reconstruction. Therefore, it is highly desirable to extract only the informative data points needed to perform a satisfying ptychographic reconstruction with minimum computational cost. 

Various techniques have been developed that use information directly extracted from diffraction patterns to approximate the desired information regarding an object. For example, in the aforementioned approach of pre-coarse scan to obtain a rough RoI for ptychography, the absorption contrast can be obtained from the total transmitted intensity at each scan point to provide a coarse and quick overview of the scanned region~\cite{thibault2008high}. For weak absorbing objects,  obtaining good absorption contrast in hard X-rays is challenging. Since phase gradients in the object can deflect X-rays and result in intensity redistribution on the detector, a few methods based on this mechanism have been proposed to visualize weak absorbing objects. For example,  scattering power evaluation was proposed to map out the ratio of scattered photons and incident photons at individual scan positions to give immediate online feedback on the location of a biological specimen~\cite{dierolf2010ptychographic}; and differential phase contrast  obtained by plotting moments of diffraction patterns as a function of positions also provides strong contrast over absorption for objects composed of light elements~\cite{hornberger_jsr_2008,deng_pnas_2015}. Beyond scattering power and moments,  more information is embedded in the diffraction patterns. For example, sample feature size, shape, and orientation can be extracted in small-angle X-ray scattering to obtain a deeper understanding of material properties~\cite{weinhausen_njp_2012, schaff2015six}.
While these techniques provide immediate feedback and mapping of scattering/transmission properties of the object within the field of view, an automatic and robust method to identify "important" data to be used for the reconstruction is lacking. Therefore, we explore machine learning techniques to achieve automatic data selection without human intervention. In particular, unsupervised learning has drawn a great attention in scientific community because of its robustness without requiring labeled data, which is often difficult to acquire because of the lack of ground truth. Unsupervised learning has been successfully applied in different aspects of ptychography, such as identifying probes with diminished quality~\cite{lin2021classification} or clustering different oxidation behavior inside materials~\cite{hirose2019oxygen}. 

In this work we utilize the combined knowledge of object transmission and scattering orientation information (in the form of the center of mass of each diffraction pattern) and propose a physics-informed unsupervised learning algorithm to identify diffraction patterns at scan positions that are within the RoI. As an outcome, only the identified diffraction patterns are used for the reconstruction, saving valuable computational resources while achieving reconstruction quality comparable to that when using the entire dataset.

\section{Methods}
We describe $K$ diffraction patterns $\mathbf{P}^{k} \in \bbR^{N\times M}$, $(k=1,\cdots,K)$, where $N\times M$ is the corresponding image size. 
The two-dimensional center of mass (CoM) array can then be calculated as

\begin{equation}
\label{eq:centerofmass}
(O_{x}^k,O_{y}^k) = \left(\frac{1}{T^{k}}\sum_{i=1}^{N}i\sum_j \mathbf{P}_{j,i}^k, \, \frac{1}{T^{k}}\sum_{j=1}^{M}j\sum_i \mathbf{P}_{j,i}^k\right),
\end{equation}
where $T^{k}$ is the sum of all pixel values over the entire image, namely, $T^k=\sum_{i,j}\mathbf{P}_{i,j}^k$.
Calculating the CoM according to \eqref{eq:centerofmass} for K diffraction patterns returns $\mathbf{O} \in \bbR^{K\times 2}$, whose row $k$ represents the $(x,y)$ coordinates of the CoM of the $k$th diffraction pattern.

To calibrate for beam-center offset errors and represent each row of $\mathbf{O}$ as a vector  having an origin in the calibrated center of image, we  transform $\mathbf{O}$ to its standard normal distribution form $\vec{\mathbf{O}}$ with zero mean and unit variance as 
\begin{equation}
\label{eq:calibrate}
\vec{\mathbf{O}}=\left[\left(\vec{O}_{x}^k,\vec{O}_{y}^{k}\right)\right]_{k=1}^K = \left[\left(\frac{O_{x}^k-\bar{O}_{x}}{\sigma_{x}},\frac{O_{y}^k-\bar{O}_{y}}{\sigma_{y}}\right)\right]_{k=1}^K, 
\end{equation}
where $(\bar{O}_{x},\bar{O}_{y})$ and $(\sigma_{x},\sigma_y)$ are the mean and variance of $\mathbf{O}$, respectively.

\begin{figure}[!ht]
\centering
\begin{tabular}{cc}	
	\includegraphics[width=0.4\linewidth]{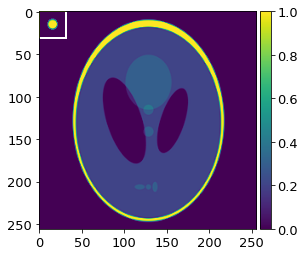} &
	\includegraphics[width=0.4\linewidth]{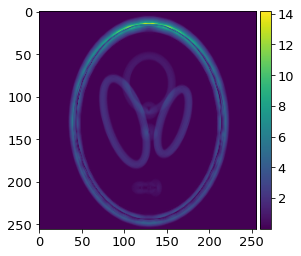} \\
	(a) & (b)\\
\end{tabular}
\caption{
(a) Shepp-Logan phantom as the object being raster scanned by a simple circular probe shown in the top left corner. (b) Corresponding magnitude of $\vec{\mathbf{O}}$ (after scanning the object using the probe), highlighting the area with feature changes, e.g., boundaries of each individual ellipse. 
}
\label{fig:simulation}
\end{figure}

The magnitude of each row vector of $\vec{\mathbf{O}}$ is correlated with the strength of the scattered light caused by the object at scanning position $k$ and yields characteristic information regarding the outline and shape of the object.
As an example, using simulated diffraction patterns obtained from using a Shepp-Logan phantom as the object and a small round probe (see Fig.~\ref{fig:simulation}(a)), we calculate and plot the magnitude of each row of $\vec{\mathbf{O}}$ at each scan position. 
As shown in Fig.~\ref{fig:simulation}(b), the magnitude metric gives us an outline of the feature changes in terms of scattering strength and can be used to extract the RoI.

\begin{figure}[!ht]
\centering
\begin{tabular}	{cc}
	\includegraphics[width=0.43\linewidth]{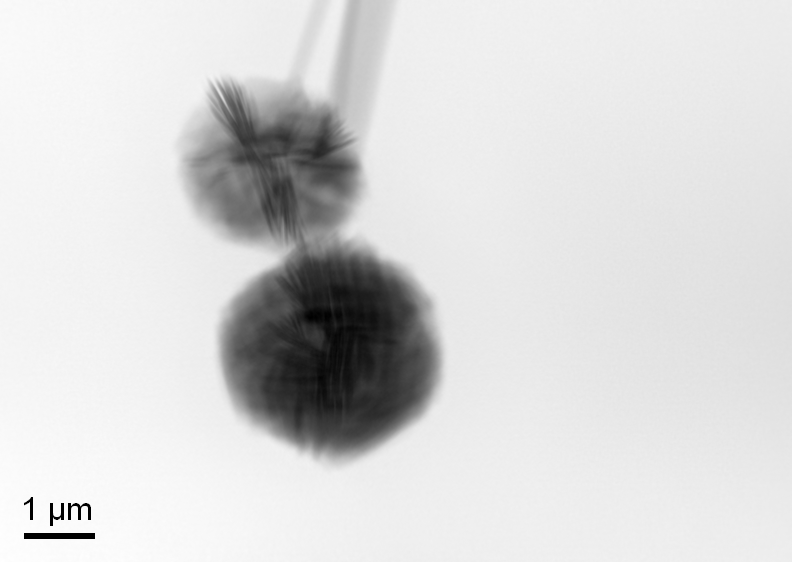} & 	\includegraphics[width=0.45\linewidth]{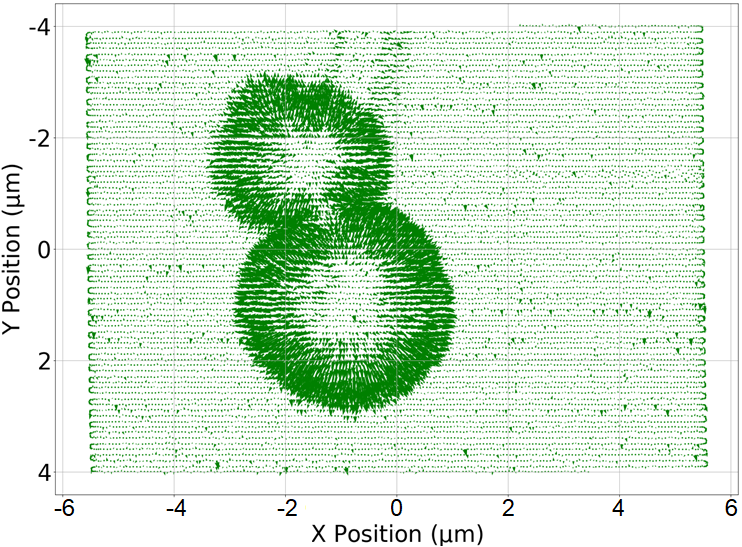} \\
	(a) &
	(b) \\
\end{tabular}
\caption{
(a) Reconstruction with all diffraction patterns. 
(b) Quiver plot of the calculated magnitude of the center of mass vector for each diffraction pattern at its scan position, which highlights the feature change area, along with the two bands at the top of the image.
}
\label{fig:algorithmWalkthroughNEW}
\end{figure}

\begin{figure}[!ht]
\centering
\begin{tabular}{cc}	
	\includegraphics[width=0.5\linewidth]{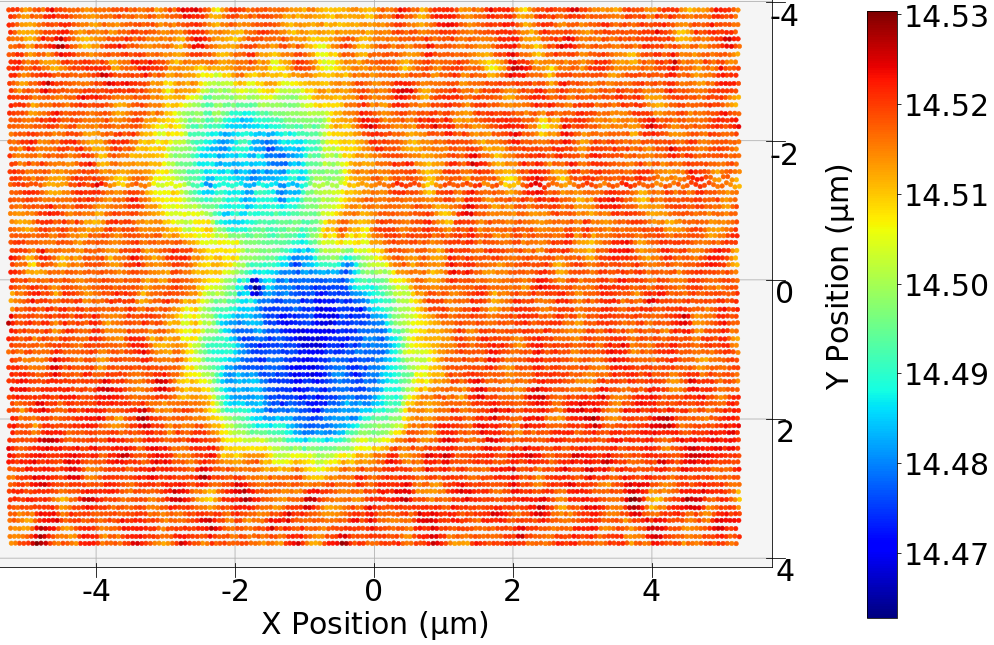} &
	\includegraphics[width=0.5\linewidth]{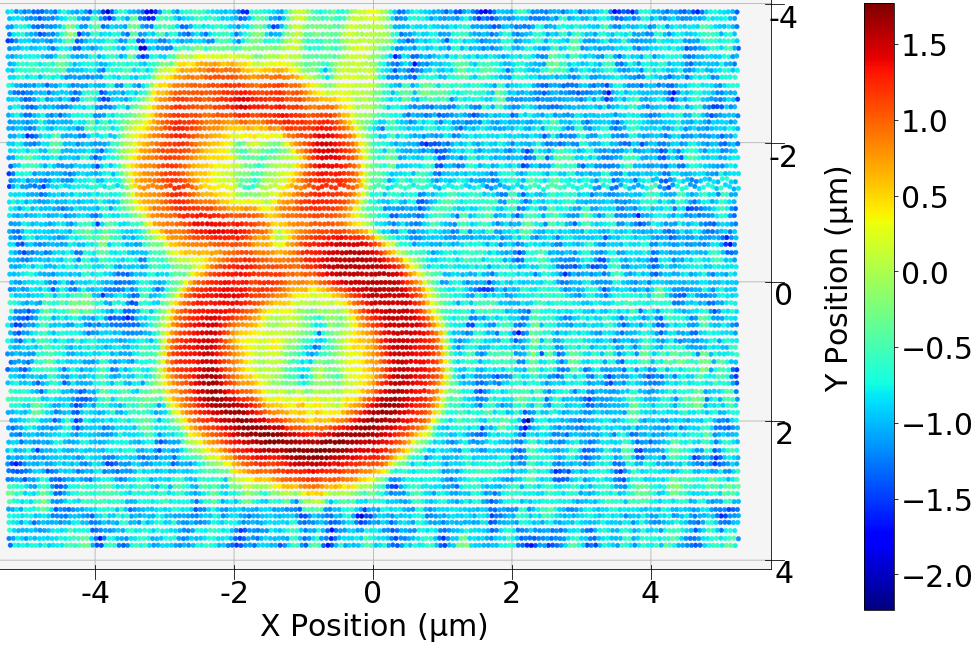}  \\
	(a) & (b) \\
\end{tabular}
\caption{
(a) Scatter plot of the log value of the sum of all pixels in each diffraction pattern at its scanning position. 
(b) Scatter plot of the log value of the magnitude of each arrow in quiver plot seen in Fig.~\ref{fig:algorithmWalkthroughNEW}(b).
}
\label{fig:algorithmWalkthrough2NEW}
\end{figure}

We further demonstrate the characteristic information embedded in the CoM array of an experimental dataset that was acquired by the Velociprobe instrument~\cite{deng2019velociprobe} at the Advanced Photon Source. 
CuS secondary particles composed of a batch of nanosheets were imaged at 8.8 keV with the fly-scan mode, in which the object moved continuously as a series of diffraction patterns acquired by an Eiger X 500K detector. 
A total of $K = 15,980$ diffraction patterns,  covering an area of 11 $\mu$m $\times$ 8 $\mu$m, were collected at a continuous frame rate of 200 Hz in less than 80 seconds. 
Each diffraction pattern was cropped to 256 $\times$ 256 and reconstructed by the generalized least-squares maximum likelihood algorithm~\cite{odstrvcil2018iterative} implemented in the PtychoShelves package~\cite{wakonig2020ptychoshelves}.
Figure \ref{fig:algorithmWalkthroughNEW}(a) shows the reconstructed phase using the complete dataset. 
The CoM vector array for this dataset is calculated by using Eq.~\ref{eq:centerofmass} and is shown in Fig.~\ref{fig:algorithmWalkthroughNEW}(b) as a quiver plot, providing a map for the scattering information of the object.
As a complementary modality to the oriented scattering information present in the CoM array, we further exploit the absorption contrast image, as obtained from the total signal recorded on the pixelated area detector at each scan position, $T^k$, which is also used when calculating the CoM vector for each diffraction pattern.
In order to filter out noisy data, a mean filter of size 3x3 is applied to both the absorption contrast image and the CoM magnitude image, which are shown in log scale in Fig.~\ref{fig:algorithmWalkthrough2NEW}(a) and Fig.~\ref{fig:algorithmWalkthrough2NEW}(b), respectively. 


\begin{figure}[!ht]
\centering
\begin{tabular}{cc}	
	\includegraphics[width=0.4\linewidth]{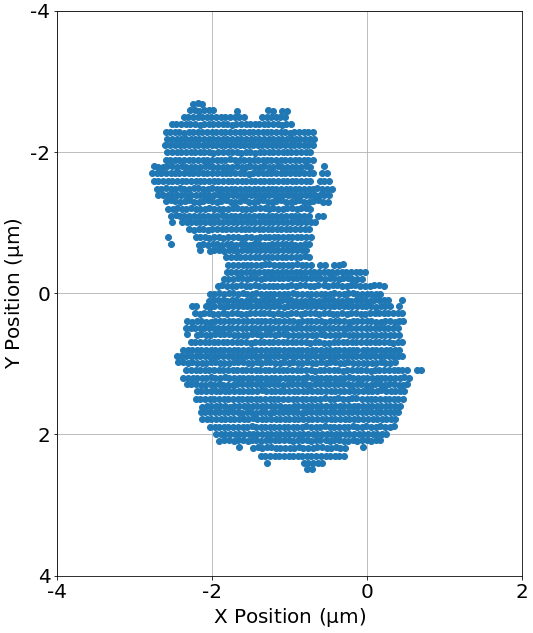} &
	\includegraphics[width=0.4\linewidth]{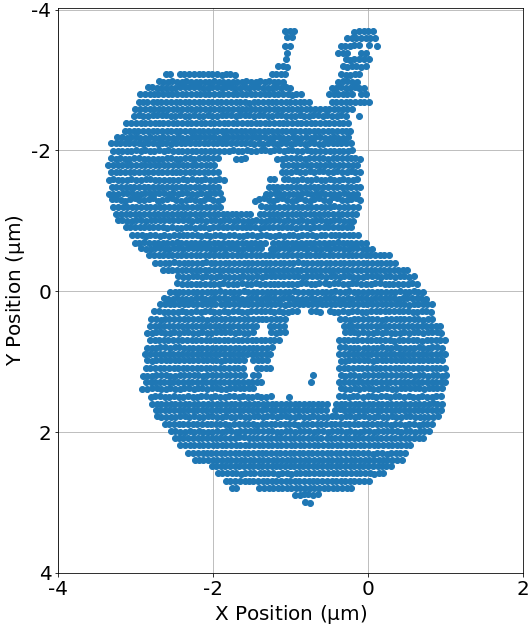} \\
	(a) & (b)\\
	\includegraphics[width=0.4\linewidth]{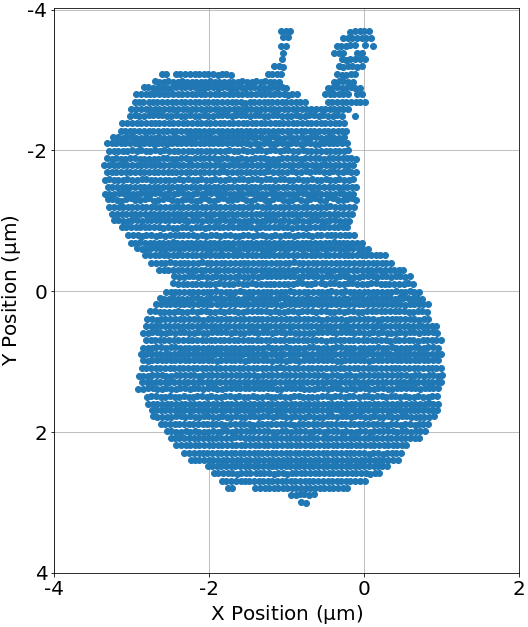} &
	\includegraphics[width=0.3\linewidth]{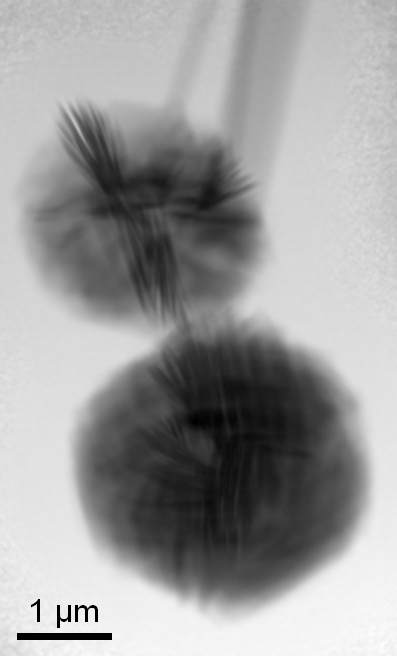} \\
	(c) & (d)\\
\end{tabular}
\caption{
(a) RoI cluster  identified as high absorption region. 
(b) RoI cluster identified as high scattering region. 
(c) Combined RoI clusters (a) and (b).
(d) Reconstruction of object using only retained diffraction patterns in (c).
}
\label{fig:algorithmWalkthrough3NEW}
\end{figure}

Given the characteristic information embedded in these two complementary image modalities, we now explore an unsupervised clustering technique to classify different types of features.
First, we apply k-means~\cite{wu2012advances} (10 iterations) to partition the absorption contrast image into two clusters.
The cluster with the lower values corresponds to diffraction patterns with lower light transmission due to the object absorption, and vice versa. 
A total of 1,832 diffraction patterns are identified and labeled as blue dots in their corresponding scan position in Fig.~\ref{fig:algorithmWalkthrough3NEW}(a), which constitutes as the first part of the RoI cluster. 
Similarly, we apply k-means to partition the CoM array magnitude image into two clusters. The cluster of diffraction patterns with higher magnitude values corresponds to higher gradient changes in terms of scattering strength and contributes to the outline of the RoI and the second part of the RoI cluster. 
In this dataset, a total of 2,998 diffraction patterns belonging to this cluster are identified and shown in Fig.~\ref{fig:algorithmWalkthrough3NEW}(b). 
Next, we combine these two identified clusters to obtain the complete RoI cluster; counting overlapping diffraction patterns only once, a total of 3,260 diffraction patterns are obtained, reducing the RoI data to 20\% of the total data, as shown in Fig.~\ref{fig:algorithmWalkthrough3NEW}(c).
The corresponding reconstruction using only the identified diffraction patterns from the RoI is shown in Fig.~\ref{fig:algorithmWalkthrough3NEW}(d).

\section{Validation of Reconstruction Quality}
Since the ultimate goal is to optimally use the computational resources only on "important" data without sacrificing the reconstruction quality, we now qualitatively examine the reconstruction quality using only the identified diffraction patterns that represent the RoI. Once the RoI is identified (in terms of the scanning positions), we can study the effect of the RoI accuracy on reconstruction quality by adjusting the size of the area (referred to as "border size") to be included in reconstruction. 

\begin{figure}[!ht]
\centering
\begin{tabular}{cc}	
	\includegraphics[width=0.4\linewidth]{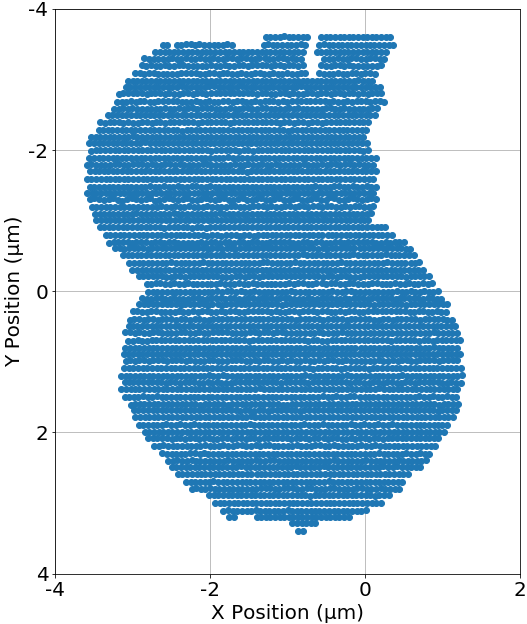} &
	\includegraphics[width=0.4\linewidth]{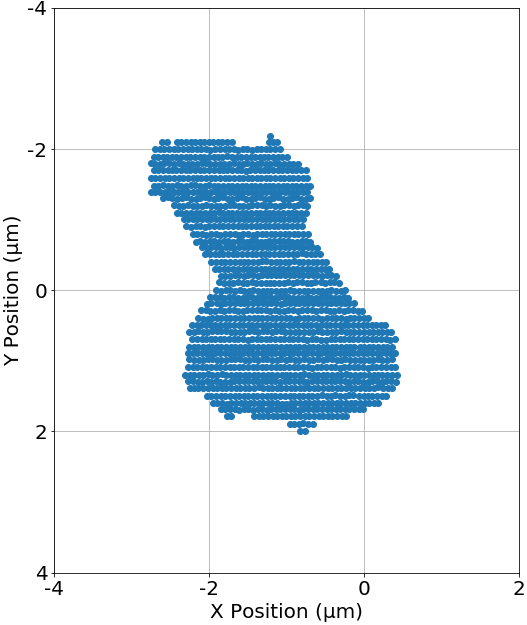} \\
	(a) & (b)\\
	\includegraphics[width=0.3\linewidth]{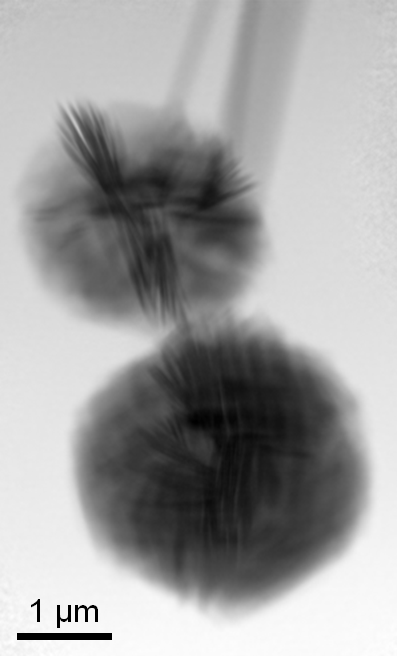} &
	\includegraphics[width=0.3\linewidth]{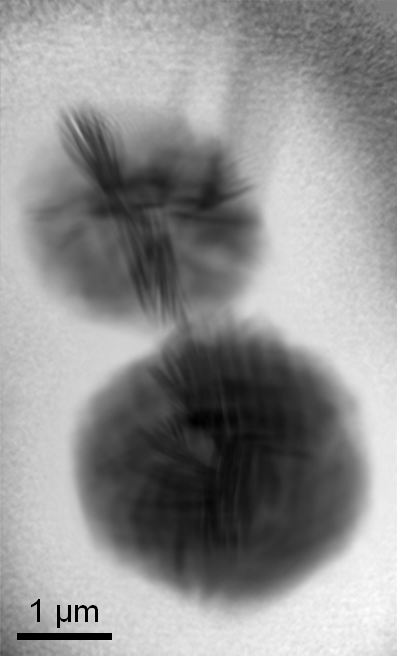} \\
	(c) & (d)\\

\end{tabular}
\caption{
Given the identified RoI shown in Fig.~\ref{fig:algorithmWalkthrough3NEW}(c), (a) \& (c): Expanding the RoI by four step sizes outwards along the x-direction and its corresponding reconstruction; (b) \& (d): Shrinking the RoI by eight step sizes inwards along the x-direction: and its corresponding reconstruction.
}
\label{fig:increaseShrinkRecon}
\end{figure}

As an example, increasing the border size in Fig.~\ref{fig:algorithmWalkthrough3NEW}(c) by four step sizes outwards along the x-direction and y-direction 
increases the RoI data from 20\% to 26\%, as shown in Fig.~\ref{fig:increaseShrinkRecon}(a). 
Similarly, decreasing the border size by eight step sizes 
in Fig.~\ref{fig:algorithmWalkthrough3NEW}(c) inwards along the x-direction and y-direction results in a further reduction of 10\% of the total diffraction patterns, as shown in Fig.~\ref{fig:increaseShrinkRecon}(b). 
The resulting reconstructions by either increasing the RoI area or decreasing the RoI area are shown in Fig.~\ref{fig:increaseShrinkRecon}(c) and Fig.~\ref{fig:increaseShrinkRecon}(d), respectively. 
We can see that the reconstructions of the true sample area are comparable; however, as expected, the boundary of the FoV shown in Fig.~\ref{fig:increaseShrinkRecon}(d) is noisier when compared with Fig.~\ref{fig:increaseShrinkRecon}(c) because of 
the lack of diffraction patterns from these locations.

To more quantitatively examine the reconstruction qualities when using only the reduced number of diffraction patterns,
we compare the reconstructions between using the full dataset and using our down-selected diffraction patterns corresponding to the identified RoI.
To be more specific, using the reconstruction from the full set of diffraction images (Fig.~\ref{fig:algorithmWalkthroughNEW}(a)) 
as the reference image, we calculate the structural similarity index measure (SSIM) \cite{wang2004image} between the reference image and other partial reconstructions.
Briefly, we use the state-of-the-art package \cite{wakonig2020ptychoshelves} for the reconstruction, while setting the batch size as 10\% of the total diffraction patterns, and we perform five independent reconstructions for each test case to obtain an average behavior.
All the numerical experiments are implemented in MATLAB and
performed on a platform with a single NVIDIA GeForce RTX 2080 GPU.

\begin{figure}[!ht]
\centering
\includegraphics[width=0.8\linewidth]{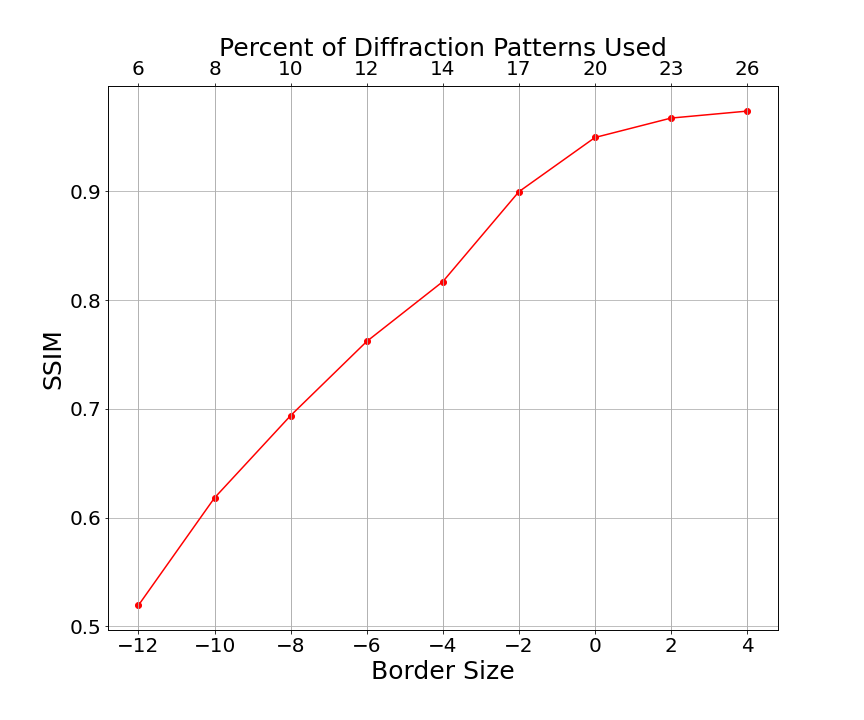}
\caption{
Reconstruction quality measured by SSIM between the reference image (reconstruction using the full dataset) and the reconstruction obtained by different RoIs (as adjusted by different border sizes). Each point is the mean value of five independent reconstructions. A higher border size gives a higher SSIM due to  more diffraction patterns surrounding  the identified RoI in the reconstruction, whereas a negative border size shrinks the RoI and includes fewer diffraction patterns, therefore yielding a lower SSIM. The identified RoI is  robust in the sense that a small adjustment does not change the reconstruction quality significantly. 
}
\label{fig:SSIMvBORDER}
\end{figure}

In Fig.~\ref{fig:SSIMvBORDER}, as we decrease the border size to shrink the area of RoI, in other words reducing the number of diffraction patterns used in the reconstruction, the reconstruction quality degrades gradually, as expected. 
Meanwhile, the SSIM plateaus as we increase the border size to around 2; in other words, increasing the number of diffraction patterns beyond this point yields marginal benefits for the reconstruction quality of the object, thus suggesting the effectiveness of our identified RoI. 
Next, we provide an empirical complexity analysis to demonstrate the overall computational cost reduction achieved by our proposed method. 
The time needed for the reconstruction with all diffraction patterns is 104 minutes, 
whereas running the reconstruction using the retained diffraction patterns (specifically, border size = 0) as identified with our method  takes only 23 minutes, which is roughly only 20\% of the time required for the full reconstruction. 
The total time needed for the preprocessing step to identify the RoI is on the order of 10 seconds, which includes the center of mass calculation and the k-means clustering.
Therefore, the preprocessing time in general is negligible when compared with the expensive reconstruction. 

\section{Discussion}
We propose a physics-inspired unsupervised learning framework to successfully classify the "important" data corresponding to the region of interest, which ultimately leads to a decrease in time for the reconstruction while preserving high reconstruction quality. More importantly, we exploit information provided by different image modalities that are correlated but also complementary. For example, the absorption contrast image is mostly effective for identifying regions with high optical density but fails in regions with low optical density.
On the other hand, the scattering orientation information, in the form of center of mass of each diffraction pattern, excels in detecting scattered light changes, which is effective in regions with feature changes. The combined use of multimodal information from both the absorption contrast image and scattering orientation image, as directly extracted from diffraction patterns, complements the otherwise incomplete object information. 

We demonstrate the application of our proposed method on an experimental dataset and show that with the identified "important" diffraction patterns, the total processing time including any preprocessing and reconstruction can be reduced to only 20\% of the original reconstruction scheme using the full dataset, without sacrificing reconstruction quality. 
Our developed method has the potential of solving a myriad of problems that may benefit from obtaining an immediate preview of the resulting reconstruction, identifying diffraction patterns that relate to the region of interest, and removing unwanted data prior to committing time to the reconstruction.
We note that our proposed method is mostly efficient for a focused beam when the used beam size is comparatively smaller than the object of interest so that contrast variation can be mapped out. For a large beam with faint absorption and scattering contrasts, a future direction will be to explore deep clustering of diffraction pattern with more clustered features rather than only two as used in this paper. 
Another interesting future direction is, instead of performing a simple raster scan over the entire field of view, to extend our proposed method to provide "important" feedback in real time and facilitate the optimal experimental scanning pattern based on the morphology of the object, in order to  save the overall data acquisition time.

\section{Acknowledgments}
This material is based upon work supported by the U.S. Department of Energy, Office of Science, under contract number DE-AC02-06CH11357.

\bibliographystyle{unsrt}  
\bibliography{references}

\end{document}